\documentstyle[aps,prb,psfig]{revtex}

\begin{document}

\setlength{\unitlength}{1.17cm}

\title{One and Two Dimensional Spin Systems in the Regime Close to
Deconfinement of Spinons}

\author{T. M. R. Byrnes, M. T. Murphy, O. P. Sushkov}
\address{School of Physics, The University of New South Wales, Sydney 2052, Australia}

\draft

\maketitle
\begin{abstract}
Based on the Majumdar-Ghosh chain we construct several spin models which
allow us to investigate spinon dynamics in the regime close to deconfinement
of spinons. We consider the $J_1-J_2-\delta$ model, two coupled $J_1-J_2$
chains (ladder), and a 2D array of coupled $J_1-J_2$ chains.
Using the picture of two spinons interacting with a string confining
potential we calculate the singlet-triplet splitting, magnetic structure 
factor, tunneling amplitude of two spinons and the excitation spectra for the
ladder and the array.
\end{abstract}

\section{Introduction}

Interest in low dimensional systems with dimerization
 has greatly increased in recent years due to several experimental findings. 
Compounds such as (VO)$_2$P$_2$O$_7$ \cite{barnesgarrett}, 
Cu(NO$_3$)$_2\cdot 2.5$H$_2$O \cite{bonner}, CuWO$_4$ \cite{lake} and 
Cu$_2$(C$_2$H$_{12}$N$_2$)$_2$Cl$_4$ \cite{chaboussant} effectively behave as 
one-dimensional chains with intrinsic dimerization. 
Further motivation for the interest comes with the discovery of a 
spin-Peierls transition in the dimerized, weakly coupled, ladder system, 
CuGeO$_3$ \cite{hase} which now has a well determined magnetic structure 
\cite{lorenzoetc}. Finally, the recent discovery that the ladder system, 
Sr$_{0.4}$Ca$_{13.6}$Cu$_{24}$O$_{41.84}$, becomes a superconductor at 12 K 
and 3 GPa \cite{greven} has renewed the idea, originally suggested by 
Anderson \cite{anderson}, that spin chain dynamics is linked with the 
mechanism of high-$T_c$ superconductivity.

The interest in systems with dimerization is not restricted to the 1D case.
There are a number of 2D theoretical models with induced or spontaneous
dimerization, such as the 
two-layer Heisenberg model (see, e.g. Ref. \cite{Chub}),
dimerized Heisenberg models \cite{Singh} and the $J_1-J_2$ model 
\cite{Sachdev}. 
Some of the models can be relevant to real materials \cite{Normand}.

A very important question relevant to any disordered
quantum spin system is: what are the elementary excitations of the system?
Until recently, common wisdom was that there are two possibilities:
1) elementary excitations have spin 1 (elementary triplets) or
2) elementary excitations have spin $1/2$ (spinons). See Ref. \cite{Sach} for a
review. However, now it is becoming clear that the spectrum of excitations of
disordered quantum spin systems is much more complicated.
The spin $1/2$ 1D Heisenberg chain is a 
very important example of a pure spinon excitation
spectrum \cite{Fad}. Some generalizations of this model, such as the frustrated
Heisenberg chain, also have pure spinon excitations. However, introduction of
external dimerization changes the spectrum drastically: the spinons become
confined and the spectrum consists of multiple singlet and triplet states
\cite{Affleck,Uhrig}. On the other hand, at very strong dimerizations, one would expect pure triplet excitations. This is, however, not the case. It has been 
demonstrated that there are multiple bound states of the triplet excitations
\cite{Dam,Sushkov,Kotov}. Moreover, the bound state can sometimes have 
energies very close to 
the ``elementary triplet'', as in the 2D $J_1-J_2$ model \cite{Kotov1}.
The complexity of the spectra of quantum spin systems is somewhat similar to the 
complexity of the meson spectra in Quantum Chromodynamics. In reality, the 
physics is also very similar. 

The purpose of the present study is to investigate the 
excitation spectra of the spin 
systems which are close to deconfinement of spinons. As a building block we use
the frustrated $J_1-J_2$ spin $1/2$ chain near the Majumdar-Ghosh point
($J_2/J_1=0.5$). Spinons are deconfined in this model. 
The models we consider have bound states of arbitrarily large size,
meaning that they are close to deconfinement.
The rest of the paper is organized as follows. In Section II we consider
a $J_1-J_2-\delta$ chain at $J_2 \approx 0.5 J_1$. Following the approaches 
suggested by  Affleck \cite{Affleck} and Uhrig et al\cite{Uhrig} we calculate
singlet-triplet splittings and spin structure factors (quasiparticle residues
for triplet excitations). In Section III we consider two coupled $J_1-J_2$
chains (a ladder). The most interesting part here is the 
tunneling of the spinon from
one leg of the ladder to another. In Section IV we consider two-dimensional
array of the $J_1-J_2$ chains. We calculate the confinement radius and
demonstrate that the low energy dispersion of the triplet excitation is isotropic.
Section V presents our conclusions.

\section{One dimensional $J_1-J_2-\delta$ model}
The Hamiltonian for this model is of the form
\begin{equation}
\label{ham}
H = J\sum_{i}\left[\left(1+(-\delta)^{i+1}\right){\bf S}_{i}{\bf S}_{i+1} 
+ \alpha{\bf S}_{i}{\bf S}_{i+2}\right]
\end{equation}
where ${\bf S}_i$ represents spin $1/2$ at the site $i$.
We set $J_1=J$, $J_2=\alpha J$ and denote the degree of dimerization as $\delta$.
Consider first the case of zero explicit dimerization $\delta=0$.
It is well known that at $\alpha < \alpha_c \approx 0.241$ 
\cite{Jullien,Okamoto} there is a unique ground state and that the excitation spectrum is
gapless. On the other hand, at $\alpha > \alpha_c$, the ground state has
spontaneous dimerization and is therefore doubly degenerate, with a gap in the excitation
spectrum.
It has been shown by Majumdar, Ghosh and van den Broek \cite{majumdar} 
that, at $\alpha=1/2$, all quantum fluctuations are canceled out and
the ground state is particularly simple:
the product of exact spin dimers,
\begin{equation}
\label{gs}
\left|0\right> =\,.\,.\,.\,\left[j-2,j-1\right]\left[j,j+1\right]
\left[j+2,j+3\right]\,.\,.\,.\,\,.
\end{equation}
where we denote the singlet pair formed from adjacent sites, $i$ and $j$, by
 $\left[i,j\right]$. The singlet chain shifted by unity with respect to 
Eq.\ (\ref{gs}) is also a valid ground state but the degeneracy is broken when 
$\delta\neq 0$. Due to the simplicity of the MG point we largely restrict our 
analysis in the present work to the case of $\alpha=\frac{1}{2}$.

Another important development was the analysis of the excitations on a MG chain by 
Shastry and Sutherland \cite{shastry}. The simplest excitation consists of a pair
of propagating topological kinks (spinons), see Fig. 1.
\begin{figure}
\label{fig:prop}
\begin{center}
\begin{picture}(5.9,1.0)
\multiput(0.4,0.4)(1.8,0){2}{\oval(0.7,0.14)}
\multiput(3.8,0.4)(1.8,0){2}{\oval(0.7,0.14)}
\multiput(1.3,0.4)(3.4,0){2}{\vector(0,1){0.42}}
\put(2.8,0.4){...}
\end{picture}
\end{center}
\renewcommand{\baselinestretch}{0.8}
\caption{Propagation of a pair of spinons (arrows mark the unpaired spins) on the MG chain. Ovals represent singlet pairs. Note that the spinons may not pass through each other due to the dimerized nature of the chain.}
\end{figure}
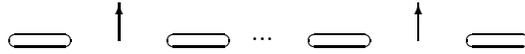
Using the wave function corresponding to Fig. 1 as a variational function one can easily find the
spinon dispersion
\begin{equation}
\label{eq:disp}
\epsilon_k = \frac{5}{8}J+\frac{1}{2}J\cos{2k}.
\end{equation}
The energy gap is therefore $\Delta=J/8$. The dispersion (\ref{eq:disp})
can be expanded near minimum, $q=k-\pi/2$, as
\begin{equation}
\label{d2}
\epsilon_q=\Delta+{{q^2}\over{2m}},
\end{equation}
with effective spinon mass $m=0.5/J$.
In spite of the variational
nature of the dispersion, (\ref{eq:disp}) agrees very well with exact numerical results \cite{sorensen,Zheng}. The scattering continuum for a pair of spinons with total
momentum $K$ is shown on Fig. 2.
\begin{figure}
\centerline{\psfig{width=8cm,angle=180,file=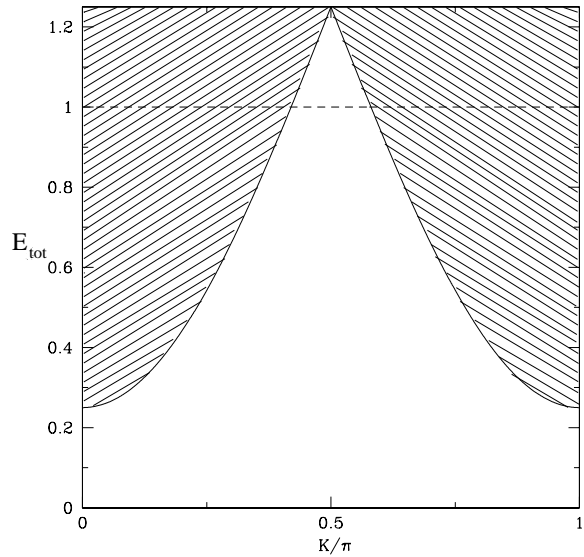}}
\label{fig:shas}
\renewcommand{\baselinestretch}{0.8}
\caption[]{The scattering continuum for a pair of kinks with total momentum $K$ as found in Ref. \cite{shastry} (hashed regions) and the dispersion for the triplet with unity spinon separation and $\alpha = \frac{1}{2}$ (dashed line). $J$ has been set to unity. We see
 that the triplet configuration is a bound state at $K=\frac{\pi}{2}$.}
\end{figure}

Two spinons with parallel spins interact with one another. There is a very simple way to find this interaction.
Let us consider the state with spinons at nearest sites, see Fig. 3.
\begin{figure}
\label{fig:hop}
\begin{center}
\begin{picture}(5.6,1.1) 
\put(4.0,0.5){\oval(0.7,0.14)}
\put(1.6,0.5){\oval(0.7,0.14)}
\put(2.5,0.5){\vector(1,0){0.42}}
\put(4.9,0.5){\vector(0,1){0.42}}
\put(5.5,0.5){\vector(0,1){0.42}}
\put(0.1,0.5){\vector(0,1){0.42}}
\put(0.7,0.5){\vector(0,1){0.42}}
\put(0.55,0){\footnotesize initial}
\put(4.3,0){\footnotesize final}
\end{picture}
\end{center}
\renewcommand{\baselinestretch}{0.8}
\caption{Triplet hopping with unity kink separation.}
\end{figure}
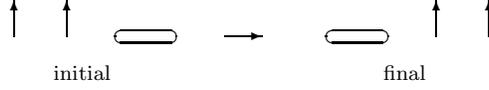
This corresponds to the triplet excitation of the spin-singlet dimer.
The excitation energy of the triplet is $J$. In general the triplet can hop to the 
nearest bond, but the hopping matrix element $0.25J(1-2\alpha)$ vanishes at the MG point. Thus, the energy of the triplet shown in Fig. 2 by the dashed line
is independent of momentum: $E_{triplet}=J$. Certainly this is not an exact
result. Quantum fluctuations ensure there are some corrections to this energy, 
but they are very small.
The triplet does not exist as a stationary state when its energy is inside the
two spinon continuum. However, we see from the Fig. 2 that there is a window
where it does exist. This state has been found earlier in Ref. \cite{shastry} 
using a more sophisticated method. 
The triplet exists because of the effective attraction between the spinons at 
nearest sites which we can write as 
\begin{equation}
\label{Veff}
V_{eff}(x_1,x_2)=C\delta_{x_1+1,x_2},
\end{equation}
where $x_1$ and $x_2$ are the spinons' coordinates which take integer values.
To find the constant $C$ let us write down wave function of the triplet with 
momentum $K$ as
\begin{equation}
\label{twf}
\Psi_K(x_1,x_2)\propto e^{iKX}\delta_{x,1} \propto e^{iKX}\sum_k e^{ik(x-1)},
\end{equation}
where $X=(x_1+x_2)/2$ is the center of mass coordinate and $x=x_2-x_1$ is 
the relative coordinate of the two spinons. In the bound triplet, $x=1$.
 Eq. (\ref{twf}) can be transformed to
\begin{equation}
\label{twf1}
\Psi_K(x_1,x_2) \propto \sum_k e^{-ik} e^{i({K\over2}+k)x_2}e^{i({K\over2}-k)x_2}.
\end{equation}
In this wave function all momenta contribute with equal weight, therefore
the average kinetic energy is
\begin{equation}
\langle \Psi| E_{kin}|\Psi \rangle = \int_{-\pi/2}^{\pi/2}{{dk}\over{\pi}}\left[ 
\epsilon\left(K/2 + k\right) + 
\epsilon\left(K/2 - k\right)\right]={5\over 4}J,
\end{equation}
where $\epsilon(k)$ is the spinon dispersion (\ref{eq:disp}). For two spinons at 
adjacent sites the average value of the effective interaction (\ref{Veff}) is 
$\langle \Psi| V_{eff}|\Psi \rangle =C$.
However, we know that the total energy of triplet is 
$J=\langle \Psi | E_{kin}|\Psi \rangle+\langle \Psi |V_{eff}|\Psi \rangle$. 
From here we find that $C=-{1\over4}J$ and hence the effective spinon-spinon 
interaction in the triplet channel ($S=1$) is
\begin{equation}
\label{Veff1}
V_{eff}(x_1,x_2)=-{1\over4}J\delta_{x_1+1,x_2}.
\end{equation}
No such potential exists for the singlet ($S=0$) state of two spinons. 

Up till now we considered zero explicit dimerization, $\delta=0$, which
results in two degenerate ground states and deconfinement of spinons.
Let us now consider very small but nonzero dimerization. The degeneracy 
between the ground states is removed and hence creation of two spinons 
creates a string with tension $3\delta J/4$. This is confinement.
For low energy excitations one can use the quadratic approximation (\ref{d2})
and hence the wave function of the relative motion of two spinons ($x=x_2-x_1$)
obeys a Schr\"{o}dinger  equation, see Refs. \cite{Affleck,Uhrig}
\begin{equation}
\label{schr}
(E-2\Delta)\psi(x) = -\frac{1}{2\mu}\frac{\partial^2}{\partial x^2}\psi(x) + 
\frac{3\delta J}{4}x\psi(x)\, ,
\end{equation}
$\mu$ is the reduced mass, $\mu=m/2$. Boundary conditions are important here.
The boundary condition at large $x$ is obvious: $\psi(\infty)=0$.
Boundary conditions for small $x$ are different for triplets and singlets
\begin{eqnarray}
\label{bound}
triplet: \ \ \ && \psi^t(-1)=0,\\
singlet: \ \ \ && \psi^s(+1)=0.\nonumber
\end{eqnarray}
The condition for the triplet reflects the fact that the spinons (kinks) cannot
penetrate through each other, see Fig. 1. This is also true for the singlet,
but in this case there is an additional condition: excited states
must be  orthogonal to the ground state. Therefore, we assert that 
$\psi_{s}(1) = 0$  since the singlet with unity kink separation is identical 
to the ground state.  
Note that the bound states have very large sizes ($x \gg 1$) and 
,hence, in the first approximation, one can replace (\ref{bound}) by $\psi(0)=0$,
see Refs. \cite{Affleck,Uhrig}. We must use (\ref{bound}) since
we intend to consider triplet-singlet splitting.

Equation (\ref{schr}) with boundary conditions (\ref{bound}) gives spectrum
\begin{equation}
\label{sp}
E_n^{(t,s)}=2\Delta-2z_nJ\left({3\over 8}\delta\right)^{2/3}\mp {3\over 4}J \delta,
\end{equation}
where $z_n$ ($n=0,1,2...$) are the zeros of the Airy function \cite{stegun}
\begin{equation}
\label{zn}
z_0=-2.388, \ \ z_1= -4.088, \ \ ... \ \ 
z_n \approx -\left[{3\over2}\pi\left(n+{3\over4}\right)\right]^{2/3} at \ \ n\gg 1.
\end{equation}
The wave function has the form
\begin{equation}
\label{psi}
\psi_n^{(t,s)}(x) = \frac{Ai\left((x\pm 1)/\xi + z_n\right)}{\sqrt{\xi}\left|Ai^{\prime}(z_n)\right|}
\end{equation}
where $\xi \equiv (3\delta/8)^{-1/3}$ is the typical size.
The signs $\pm$ in Eqs. (\ref{sp}) and (\ref{psi}) corresponds to triplet and singlet
states. The upper sign always corresponds to the triplet.

One may conclude from Eq. (\ref{sp}) that the singlet-triplet splitting is ${3\over2}\delta J$.
However, this is not quite correct because Eq. (\ref{schr}) does not take account of
the attraction (\ref{Veff1}) in the triplet channel. This attraction decreases the triplet
energy by 
\begin{equation}
\label{spl}
\langle \psi^t|V_{eff}|\psi^t\rangle= -\frac{1}{4}
\left|\psi^t(1)\right|^{2}
=-{3\over8}\delta.
\end{equation}
Together with (\ref{sp}) this gives the singlet-triplet splitting
\begin{equation}
\label{spl1}
E^s_n-E^t_n={15\over8}\delta J
\end{equation}
which is independent of $n$.
Certainly this is not an exact result because some small quantum fluctuations were
neglected in our derivation. Nevertheless it agrees very well with numerical data 
obtained recently by S{\o}rensen {\it et al} \cite{sorensen,sorensen2}.
The data are presented in Table I.
\begin{table}
\label{tab:sor}
\begin{center}
\begin{tabular}{cccc}
$\delta$ & $E^s_n-E^t_n$ & Technique & No. of sites \\ \hline
0.050 & 0.092 & (a) & 512\\
0.025 & 0.049 & (a) & 512\\
0.005 & 0.011 & (b) & 28 \\
\end{tabular}
\end{center}
\caption{The value of the singlet-triplet splitting obtained via (a) DMRG techniques described in Ref. \cite{sorensen} and (b) exact diagonalization ($J$ has been set to unity). The finite size effects are negligible. Data provided by Erik S. S{\o}rensen.}
\normalsize
\end{table}
Our result (\ref{spl1}) is valid for small momenta of the bound states, $K\approx 0$
(or $K \approx\pi $).
Nevertheless it practically coincides with splitting which is known at $K=\pi/2$.
The singlet and triplet excited states at $\delta=0$ and $K=\pi/2$ are known
exactly \cite{caspers}.  One can show \cite{sorensen3} that these excited states remain 
eigenstates when we introduce explicit dimerization and that the triplet and singlet have 
energies $(1+\delta)J$ and $(1+3\delta)J$, and so the splitting is $2\delta J$.

In conclusion of this section let us the discuss magnetic structure factor or spectral weight
of the triplet excitations. The external perturbation is
\begin{equation}
\label{pert}
{1\over{\sqrt{2}}}\sum_x e^{iKx}S_{+x},
\end{equation}
where $S_{+x}$ is the spin raising operator at the site $x$. Then the amplitude of the
spin triplet creation at a spontaneously dimerized bond is $\left(\sin{K\over2}\right)$
\  \cite{norm}.
However this triplet is only a virtual state. To find spectral weight
of the stationary state $\psi^t_n$ we have to project the virtual state onto the
stationary one. This is proportional to the amplitude for two spinons to be at neighbouring 
sites. Hence
\begin{equation}
\label{Zn}
Z_n=\sin^2{K\over2}\left|\psi_n^t(1)\right|^2={4\over{\xi^3}}\sin^2{K\over2}=
{3\over2}\delta \sin^2{K\over2}.
\end{equation}
Our consideration of the bound states is valid only at small $K$ or at $K\approx\pi$
and hence (\ref{Zn}) is valid only at these $K$. The spectrum of triplet excitations
at $K=\pi$ is shown in Fig. 4.
\begin{figure}
\centerline{\psfig{width=8cm,angle=270,file=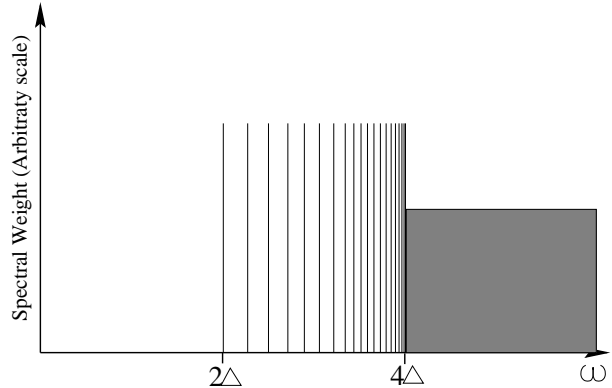}}
\label{fig:spec}
\renewcommand{\baselinestretch}{0.8}
\caption[]{Schematic plot of the spectrum of excitations described by Eqs. (\ref{sp}),(\ref{zn}),(\ref{Zn}) and (\ref{dn}). We note the gap $0 < \omega < 2\Delta$, the set of discrete bound states for $2\Delta < \omega < 4\Delta$ with spectral weight independent of $n$ and the continuum for $\omega > 4\Delta$.}
\end{figure}
 There are no excitations at $\omega < 2\Delta=0.25J$.
At $\omega > 4\Delta =0.5J$ the triplet excitation can decay into pairs (triplet+triplet
or singlet+triplet) and therefore the spectrum is continuous.
At $2\Delta < \omega < 4 \Delta$ there are a number of discrete bound states. The
spectral weight of any of these states is given by Eq. (\ref{Zn}).
The number of states, $n_{max}$, depends on $\delta$, 
such that for a small $ \delta $ there is a large $n_{max}$.
One can easily find $n_{max}$ from Eq. (\ref{sp}). For example,
at $\delta > 0.14$ there are only two states: $n=0$ and $n=1$.
At $\delta=0.01$ the number is $n_{max}=6$, and at $\delta=0.001$ the
number is $n_{max}=30$. At extremely small $\delta$ the asymptotic
formula is valid: $n_{max} \approx 0.025/\delta$. Finally the splitting
between the nearest peaks at $n \gg 1$ according to (\ref{sp}) is equal to
\begin{equation}
\label{dn}
{{dE}\over{dn}}={{1.95J\delta^{2/3}}\over{(n+3/4)^{1/3}}}.
\end{equation}

The spectral weight in Eq. (\ref{Zn}) is not quite consistent with that obtained using a series expansion method \cite{Zheng}. A possible reason for this is that convergence of the series may be very slow due to the dense spectrum (see Fig. (\ref{fig:spec})). 

\section{Two coupled Majumdar-Ghosh chains (the ladder)}
Let us consider two MG chains coupled to the ladder
\begin{equation}
\label{ham1}
H = J\sum_{i}\left\{\left[{\bf S}_{i}{\bf S}_{i+1} + 0.5{\bf S}_{i}{\bf S}_{i+2}\right]
+\left[{\bf S}_{i}^{\prime}{\bf S}_{i+1}^{\prime} + 
0.5{\bf S}_{i}^{\prime}{\bf S}_{i+2}^{\prime}\right]\right\}
+J_{\perp}\sum_i{\bf S}_i{\bf S}_i^{\prime}
\end{equation}
with $J_{\perp}\ll J$. At $J_{\perp}=0$ there are four degenerate ground states. Two of them
are topologically different and are shown in Figs. 5a and 5b. 
\setlength{\unitlength}{0.71cm}
\begin{figure}
\begin{center}
\begin{picture}(8.0,6.0)
\put(0.5,4.0){\oval(1.1,0.2)}
\put(2.5,4.0){\oval(1.1,0.2)}
\put(4.5,4.0){\oval(1.1,0.2)}
\put(6.5,4.0){\oval(1.1,0.2)}
\put(0.5,4.8){\oval(1.1,0.2)}
\put(2.5,4.8){\oval(1.1,0.2)}
\put(4.5,4.8){\oval(1.1,0.2)}
\put(6.5,4.8){\oval(1.1,0.2)}
\put(3.5,3.0){(a)}
\put(1.5,1.0){\oval(1.1,0.2)}
\put(3.5,1.0){\oval(1.1,0.2)}
\put(5.5,1.0){\oval(1.1,0.2)}
\put(7.5,1.0){\oval(1.1,0.2)}
\put(0.5,1.8){\oval(1.1,0.2)}
\put(2.5,1.8){\oval(1.1,0.2)}
\put(4.5,1.8){\oval(1.1,0.2)}
\put(6.5,1.8){\oval(1.1,0.2)}
\put(3.5,0.0){(b)}
\end{picture}
\end{center}
\renewcommand{\baselinestretch}{0.8}
\caption[]{{(a) Two chains in the `normal' state and 
(b) the `alternate' state. }}
\end{figure}
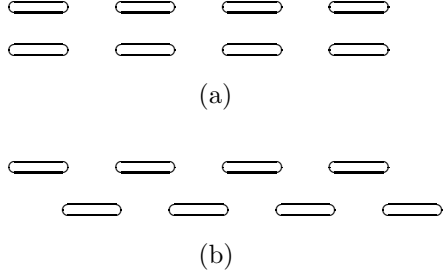
These will be called 
``normal'' and ``alternate'' states respectively. 
At $J_{\perp}\ne 0$ the degeneracy is lifted and the ``normal'' state 
corresponding to Fig. 5a becomes the true ground state. To see this,
one needs to consider the second order (in $J_{\perp}$) correction to the ground state energy.
To calculate this correction it is convenient to use a localized triplet representation for
the virtual excitation. Virtual admixtures of two triplet excitations to the ``normal''
state are shown in Fig. 6. 
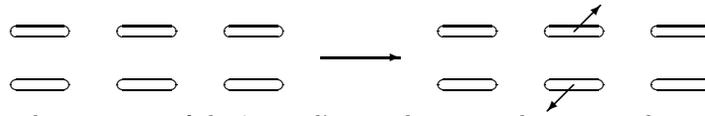
\begin{figure}
\begin{center}
\begin{picture}(15,2.0)
\put(1.5,0.0){\oval(1.1,0.2)}
\put(3.5,0.0){\oval(1.1,0.2)}
\put(5.5,0.0){\oval(1.1,0.2)}
\put(1.5,1.0){\oval(1.1,0.2)}
\put(3.5,1.0){\oval(1.1,0.2)}
\put(5.5,1.0){\oval(1.1,0.2)}
\put(6.75, 0.5){\vector(1,0){1.5}}
\put(9.5,0.0){\oval(1.1,0.2)}
\put(11.5,0.0){\oval(1.1,0.2)}
\put(13.5,0.0){\oval(1.1,0.2)}
\put(9.5,1.0){\oval(1.1,0.2)}
\put(11.5,1.0){\oval(1.1,0.2)}
\put(13.5,1.0){\oval(1.1,0.2)}
\put(11.5,0.0){\vector(-1,-1){0.5}}
\put(11.5,1.0){\vector(1,1){0.5}}
\end{picture}
\end{center}
\renewcommand{\baselinestretch}{0.8}
\caption[]{{Virtual excitations of the `normal' ground state to the 
state with two triplet pairs.}}
\end{figure}
Keeping in mind that each localized triplet has energy $J$, we
can easily find the correction $\Delta E_{normal}=-3J_{\perp}^2/8J$  \cite{cor}.
Quite similarly the correction for the ``alternate'' state due to virtual excitations shown on 
Fig. 7 gives $\Delta E_{alterate}=-3J_{\perp}^2/16J$. 
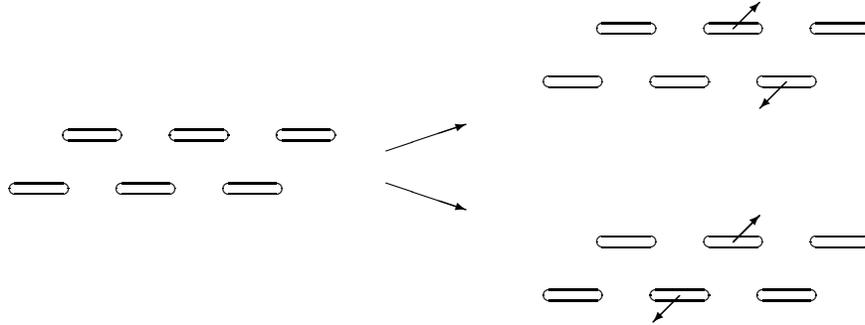
\begin{figure}
\begin{center}
\begin{picture}(18,6.0)
\put(1.5,2.0){\oval(1.1,0.2)}
\put(3.5,2.0){\oval(1.1,0.2)}
\put(5.5,2.0){\oval(1.1,0.2)}
\put(2.5,3.0){\oval(1.1,0.2)}
\put(4.5,3.0){\oval(1.1,0.2)}
\put(6.5,3.0){\oval(1.1,0.2)}
\put(8.0, 2.7){\vector(3,1){1.5}}
\put(8.0, 2.1){\vector(3,-1){1.5}}
\put(11.5,0.0){\oval(1.1,0.2)}
\put(13.5,0.0){\oval(1.1,0.2)}
\put(15.5,0.0){\oval(1.1,0.2)}
\put(12.5,1.0){\oval(1.1,0.2)}
\put(14.5,1.0){\oval(1.1,0.2)}
\put(16.5,1.0){\oval(1.1,0.2)}
\put(14.5,1.0){\vector(1,1){0.5}}
\put(13.5,0.0){\vector(-1,-1){0.5}}
\put(11.5,4.0){\oval(1.1,0.2)}
\put(13.5,4.0){\oval(1.1,0.2)}
\put(15.5,4.0){\oval(1.1,0.2)}
\put(12.5,5.0){\oval(1.1,0.2)}
\put(14.5,5.0){\oval(1.1,0.2)}
\put(16.5,5.0){\oval(1.1,0.2)}
\put(14.5,5.0){\vector(1,1){0.5}}
\put(15.5,4.0){\vector(-1,-1){0.5}}
\end{picture}
\end{center}
\renewcommand{\baselinestretch}{0.8}
\caption[]{{Virtual excitations of the `alternate' ground state to two 
different states, each with two triplet pairs.}} 
\end{figure}
Thus, the ``normal'' state has a lower
energy with the difference
\begin{equation}
\label{dE}
\Delta E= \Delta E_{alternate}-\Delta E_{normal}={{3J_{\perp}^2}\over{16J}}
\end{equation}
per two rungs along the ladder. We should note that this calculation
is not exact. Because of quantum fluctuations in the virtual states there is a small
correction to Eq. (\ref{dE}) which is also proportional to $J_{\perp}^2$. Similar to the
case of spinon dispersion (\ref{eq:disp}), we neglect this correction.

Two spinon excitations on the one leg is shown in Fig. 8. 
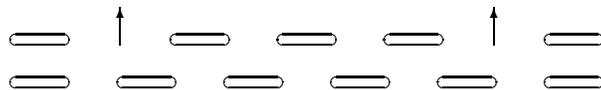
\begin{figure}[b]
\begin{center}
\begin{picture}(11,2.0)
\put(0.6,0.0){\oval(1.1,0.2)}
\put(2.6,0.0){\oval(1.1,0.2)}
\put(4.6,0.0){\oval(1.1,0.2)}
\put(6.6,0.0){\oval(1.1,0.2)}
\put(8.6,0.0){\oval(1.1,0.2)}
\put(10.6,0.0){\oval(1.1,0.2)}
\put(0.6,0.8){\oval(1.1,0.2)}
\put(2.1,0.7){\vector(0,1){0.7}}
\put(3.6,0.8){\oval(1.1,0.2)}
\put(5.6,0.8){\oval(1.1,0.2)}
\put(7.6,0.8){\oval(1.1,0.2)}
\put(9.1,0.7){\vector(0,1){0.7}}
\put(10.6,0.8){\oval(1.1,0.2)}
\end{picture}
\end{center}
\renewcommand{\baselinestretch}{0.8}
\caption[]{{Two spinon excitations on one leg of the ladder.}}
\end{figure}
It is clear from (\ref{dE})
that the tension in the confining string is $T=3J_{\perp}^2/32J$ and, therefore, the system
is described by Schr\"{o}dinger  equation (\ref{schr}) with substitution
\begin{eqnarray}
\label{subs}
&&{{3\delta J}\over{4}} \to {3\over 2}{{\left(J_{\perp}/4\right)^2}\over{J}},\\ 
&&\xi \to \left({4\over3}\right)^{1/3}\left({{4J}\over{J_{\perp}}}\right)^{2/3}.\nonumber
\end{eqnarray}
According to Eq. (\ref{psi}) the wave function for two spinons with total momentum $K$
(along the ladder) is of the form
\begin{equation}
\label{psiAB}
\psi_{nL}^{(t,s)}(x_1,x_2) = {{e^{iKX}}\over{\sqrt{N}}}
\frac{Ai\left((x\pm 1)/\xi + z_n\right)}{\sqrt{\xi}\left|Ai^{\prime}(z_n)\right|},
\end{equation}
where $X=(x_1+x_2)/2$, $x=x_2-x_1$, and the index $L=1,2$ describes the leg on which
the spinons are moving: $L=1$ corresponds to the upper leg and $L=2$ corresponds to
the lower one.
Let us consider now the specific case of triplet excitation. There is a possibility for the 
triplet to tunnel from one leg to another. The mechanism of the tunneling is shown in Fig. 9:
First the two spinons must come to adjacent sites, then the perturbation, 
$J_{\perp}{\bf S}{\bf S}^{\prime}$, swaps the triplet from one leg to another, and then
the spinons propagate along the second leg. 
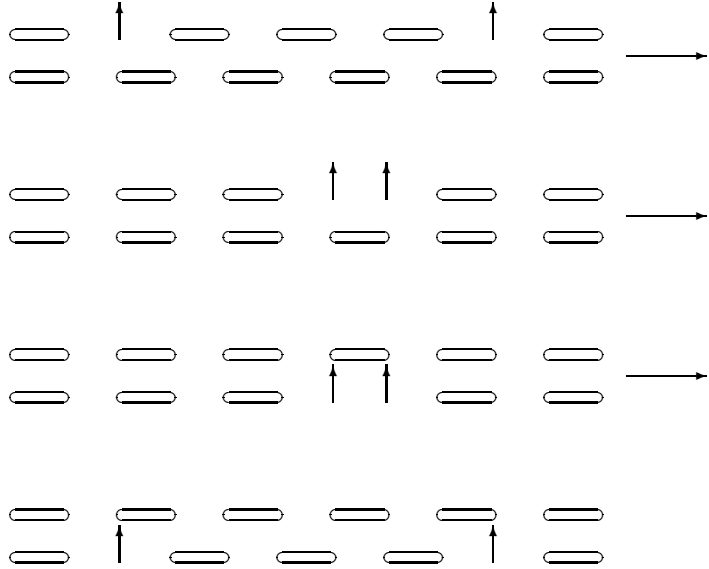
\begin{figure}[b]
\begin{center}
\begin{picture}(13,10.0)
\put(1.5,9.0){\oval(1.1,0.2)}
\put(3.5,9.0){\oval(1.1,0.2)}
\put(5.5,9.0){\oval(1.1,0.2)}
\put(7.5,9.0){\oval(1.1,0.2)}
\put(9.5,9.0){\oval(1.1,0.2)}
\put(11.5,9.0){\oval(1.1,0.2)}
\put(1.5,9.8){\oval(1.1,0.2)}
\put(3.0,9.7){\vector(0,1){0.7}}
\put(4.5,9.8){\oval(1.1,0.2)}
\put(6.5,9.8){\oval(1.1,0.2)}
\put(8.5,9.8){\oval(1.1,0.2)}
\put(10.0,9.7){\vector(0,1){0.7}}
\put(11.5,9.8){\oval(1.1,0.2)}
\put(12.5,9.4){\vector(1,0){1.5}}
\put(1.5,6.0){\oval(1.1,0.2)}
\put(3.5,6.0){\oval(1.1,0.2)}
\put(5.5,6.0){\oval(1.1,0.2)}
\put(7.5,6.0){\oval(1.1,0.2)}
\put(9.5,6.0){\oval(1.1,0.2)}
\put(11.5,6.0){\oval(1.1,0.2)}
\put(1.5,6.8){\oval(1.1,0.2)}
\put(3.5,6.8){\oval(1.1,0.2)}
\put(5.5,6.8){\oval(1.1,0.2)}
\put(7.0,6.7){\vector(0,1){0.7}}
\put(8.0,6.7){\vector(0,1){0.7}}
\put(9.5,6.8){\oval(1.1,0.2)}
\put(11.5,6.8){\oval(1.1,0.2)}
\put(12.5,6.4){\vector(1,0){1.5}}
\put(1.5,3.8){\oval(1.1,0.2)}
\put(3.5,3.8){\oval(1.1,0.2)}
\put(5.5,3.8){\oval(1.1,0.2)}
\put(7.5,3.8){\oval(1.1,0.2)}
\put(9.5,3.8){\oval(1.1,0.2)}
\put(11.5,3.8){\oval(1.1,0.2)}
\put(1.5,3.0){\oval(1.1,0.2)}
\put(3.5,3.0){\oval(1.1,0.2)}
\put(5.5,3.0){\oval(1.1,0.2)}
\put(7.0,2.9){\vector(0,1){0.7}}
\put(8.0,2.9){\vector(0,1){0.7}}
\put(9.5,3.0){\oval(1.1,0.2)}
\put(11.5,3.0){\oval(1.1,0.2)}
\put(12.5,3.4){\vector(1,0){1.5}}
\put(1.5,0.8){\oval(1.1,0.2)}
\put(3.5,0.8){\oval(1.1,0.2)}
\put(5.5,0.8){\oval(1.1,0.2)}
\put(7.5,0.8){\oval(1.1,0.2)}
\put(9.5,0.8){\oval(1.1,0.2)}
\put(11.5,0.8){\oval(1.1,0.2)}
\put(1.5,0.0){\oval(1.1,0.2)}
\put(4.5,0.0){\oval(1.1,0.2)}
\put(3.0,-0.1){\vector(0,1){0.7}}
\put(6.5,0.0){\oval(1.1,0.2)}
\put(8.5,0.0){\oval(1.1,0.2)}
\put(10.0,-0.1){\vector(0,1){0.7}}
\put(11.5,0.0){\oval(1.1,0.2)}
\end{picture}
\end{center}
\renewcommand{\baselinestretch}{0.8}
\caption[]{{The mechanism for tunnelling of two spinons. The spinons on the top leg of the ladder must first come together before they can tunnel to the lower leg. After tunneling, the spinons are free to propagate along the lower leg.  }} 
\end{figure}
Simple calculation shows that the tunneling
matrix element is equal to
\begin{equation}
\label{tun}
\tau=\langle \psi_{m2}^t|\sum_i{\bf S}_i{\bf S}_i^{\prime}|\psi_{n1}^t\rangle
={{2J_{\perp}}\over{\xi^3}}=6
{{\left(J_{\perp}/4\right)^3}\over{J^2}}.
\end{equation}
It is interesting that this is independent of the indexes $n$ and $m$.
According to (\ref{sp}) (with substitution (\ref{subs})) the energy splittings
between states with different $n$ is $\propto J_{\perp}^{4/3}$. The tunneling matrix element, 
$\tau \propto J_{\perp}^3$, is much smaller and therefore mixing within one $n$-level
is independent of the others. Effectively, at each $n$ we have a two degenerate level system 
with tunneling $\tau$ between them. The stationary states have a definite symmetry with respect 
to the legs permutation ($u$-state and $g$-state), and the triplet spectrum is
\begin{equation}
\label{sp2}
E_{n(g,u)}^t=2\Delta-2z_nJ\left({3\over4}\right)^{2/3}
\left({{J_{\perp}/4}\over{J}}\right)^{4/3}
- {{3J_{\perp}^2}\over{32J}}\pm {{3J_{\perp}^3}\over{32J^2}}+{{K^2}\over{2M}}.
\end{equation}
Here $M=2m=1/J$ is total mass of triplet excitation.

The tunneling probability 
of the singlets from one leg to another is much smaller than that for the triplets.
This is because the spinons with total spin zero cannot approach one each other: 
$\psi^s(1)=0$ (see discussion in the previous section). Neglecting this tunneling we 
can say that the singlet $g$- and $u$-states are degenerate, and that the energies of the
singlet states are given by (\ref{sp}) with substitution (\ref{subs}).

In conclusion to this section we would like to note that, for the ladder, the topological excitation with total spin 0 or 1 is possible, as shown in Fig. 10.
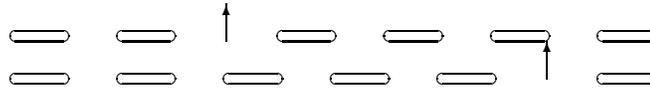
\begin{figure}
\begin{center}
\begin{picture}(13,2.0)
\put(1.5,0.0){\oval(1.1,0.2)}
\put(3.5,0.0){\oval(1.1,0.2)}
\put(5.5,0.0){\oval(1.1,0.2)}
\put(7.5,0.0){\oval(1.1,0.2)}
\put(9.5,0.0){\oval(1.1,0.2)}
\put(11.0,0.0){\vector(0,1){0.7}}
\put(12.5,0.0){\oval(1.1,0.2)}
\put(1.5,0.8){\oval(1.1,0.2)}
\put(3.5,0.8){\oval(1.1,0.2)}
\put(5.0,0.7){\vector(0,1){0.7}}
\put(6.5,0.8){\oval(1.1,0.2)}
\put(8.5,0.8){\oval(1.1,0.2)}
\put(10.5,0.8){\oval(1.1,0.2)}
\put(12.5,0.8){\oval(1.1,0.2)}
\end{picture}
\end{center}
\renewcommand{\baselinestretch}{0.8}
\caption[]{{Topological excitations with spins 1 or 0.  }}
\end{figure}
However, the gap for this excitation is $2\Delta$ and hence the threshold for
creation of two topological excitations (minimum is two) is $4\Delta$.
In the present paper we do not consider excitations with such high energies.

\section{2D array of coupled Majumdar-Ghosh chains}

Another way of linking together spin chains is to form a two-dimensional array as
shown in Fig. 11. 
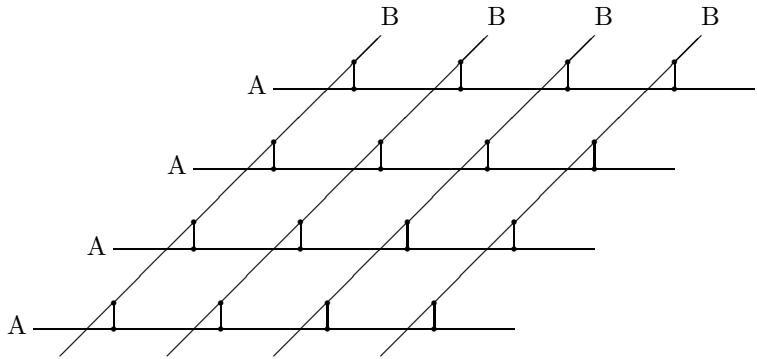
\begin{figure}
\begin{center}
\begin{picture}(13,7.0)
\put(1.0,1.0){\line(1,0){9}}
\put(0.5,0.9){A}
\put(2.5,2.5){\line(1,0){9}}
\put(2.0,2.4){A}
\put(4.0,4.0){\line(1,0){9}}
\put(3.5,3.9){A}
\put(5.5,5.5){\line(1,0){9}}
\put(5.0,5.4){A}
\put(1.5,0.5){\line(1,1){6}}
\put(7.5,6.7){B}
\put(3.5,0.5){\line(1,1){6}}
\put(9.5,6.7){B}
\put(5.5,0.5){\line(1,1){6}}
\put(11.5,6.7){B}
\put(7.5,0.5){\line(1,1){6}}
\put(13.5,6.7){B}
\multiput(2.5,1.0)(2,0){4}{\line(0,1){0.5}}
\multiput(4.0,2.5)(2,0){4}{\line(0,1){0.5}}
\multiput(5.5,4.0)(2,0){4}{\line(0,1){0.5}}
\multiput(7.0,5.5)(2,0){4}{\line(0,1){0.5}}
\multiput(2.5,1.0)(2,0){4}{\circle*{0.1}}
\multiput(4.0,2.5)(2,0){4}{\circle*{0.1}}
\multiput(5.5,4.0)(2,0){4}{\circle*{0.1}}
\multiput(7.0,5.5)(2,0){4}{\circle*{0.1}}
\multiput(2.5,1.5)(2,0){4}{\circle*{0.1}}
\multiput(4.0,3.0)(2,0){4}{\circle*{0.1}}
\multiput(5.5,4.5)(2,0){4}{\circle*{0.1}}
\multiput(7.0,6.0)(2,0){4}{\circle*{0.1}}
\end{picture}
\end{center}
\renewcommand{\baselinestretch}{0.8}
\caption[]{{The two-dimensional array. Each of the `A' and `B' chains represent spin chains similar to those previously considered. The short lines represent cross links with an exchange $ J_\perp $. }}
\end{figure}
There are a total of $N^2$ cross links. The Hamiltonian is of the
form
\begin{eqnarray}
\label{hamAB}
H &=& J\sum_x \left({\bf S}_{x,y}^A{\bf S}_{x+1,y}^A + 
0.5{\bf S}_{x,y}^A{\bf S}_{x+2,y}^A\right)
\nonumber\\
&+& J\sum_y \left({\bf S}_{x,y}^B{\bf S}_{x,y+1}^B + 
0.5{\bf S}_{x,y}^B{\bf S}_{x,y+2}^B\right)\\
&+&J_{\perp}\sum_{x,y}{\bf S}_{x,y}^A{\bf S}_{x,y}^B.\nonumber
\end{eqnarray}
The ``A''-chains are aligned along the x-direction and the ``B''-chains are aligned 
along the y-direction. 
First we want to demonstrate that the ground state of the system looks like that
shown in Fig. 12.
\begin{figure}
\begin{center}
\begin{picture}(7,5)
\multiput(0.5,0.0)(2,0){3}{\oval(1.1,0.2)}
\multiput(0.5,1.0)(2,0){3}{\oval(1.1,0.2)}
\multiput(0.5,2.0)(2,0){3}{\oval(1.1,0.2)}
\multiput(0.5,3.0)(2,0){3}{\oval(1.1,0.2)}
\multiput(0.5,4.0)(2,0){3}{\oval(1.1,0.2)}
\multiput(0.5,5.0)(2,0){3}{\oval(1.1,0.2)}
\multiput(0.0,0.5)(0,2){3}{\oval(0.2,1.1)}
\multiput(1.0,0.5)(0,2){3}{\oval(0.2,1.1)}
\multiput(2.0,0.5)(0,2){3}{\oval(0.2,1.1)}
\multiput(3.0,0.5)(0,2){3}{\oval(0.2,1.1)}
\multiput(4.0,0.5)(0,2){3}{\oval(0.2,1.1)}
\multiput(5.0,0.5)(0,2){3}{\oval(0.2,1.1)}
\multiput(0.0,0.0)(1,0){6}{\circle*{0.1}}
\multiput(0.0,1.0)(1,0){6}{\circle*{0.1}}
\multiput(0.0,2.0)(1,0){6}{\circle*{0.1}}
\multiput(0.0,3.0)(1,0){6}{\circle*{0.1}}
\multiput(0.0,4.0)(1,0){6}{\circle*{0.1}}
\multiput(0.0,5.0)(1,0){6}{\circle*{0.1}}
\end{picture}
\end{center}
\renewcommand{\baselinestretch}{0.8}
\caption[]{{The two-dimensional array. The horizontal chains are the 
A-chains while the vertical chains are B-chains. The dots represent the cross-links, the same as those in Fig. 11.}}
\end{figure}
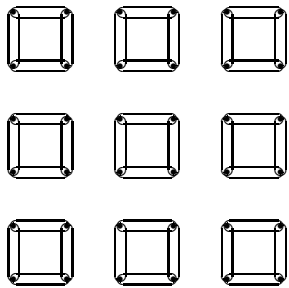
We stress that any dot in Fig. 12 denotes two spins 1/2 separated
in the vertical direction (see Fig. 11) and coupled via $J_{\perp}$.
The dimers aligned in $x$-direction are built 
from A-chain spins and the dimers aligned in $y$-direction are built from B-chain spins.
The question is: why is the ground state pattern like that in Fig. 13a
(which is actually part of Fig. 12), but not like that in Fig. 13b? 
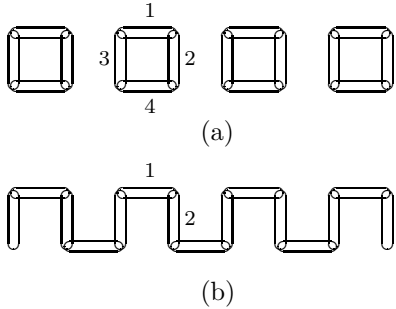
\begin{figure}
\begin{center}
\begin{picture}(7,6)
\multiput(0.5,4.0)(2,0){4}{\oval(1.1,0.2)}
\multiput(0.5,5.0)(2,0){4}{\oval(1.1,0.2)}
\multiput(0.0,4.5)(0,2){1}{\oval(0.2,1.1)}
\multiput(1.0,4.5)(0,2){1}{\oval(0.2,1.1)}
\multiput(2.0,4.5)(0,2){1}{\oval(0.2,1.1)}
\multiput(3.0,4.5)(0,2){1}{\oval(0.2,1.1)}
\multiput(4.0,4.5)(0,2){1}{\oval(0.2,1.1)}
\multiput(5.0,4.5)(0,2){1}{\oval(0.2,1.1)}
\multiput(6.0,4.5)(0,2){1}{\oval(0.2,1.1)}
\multiput(7.0,4.5)(0,2){1}{\oval(0.2,1.1)}
\put(2.45, 3.5){{\footnotesize $ 4 $}}
\put(3.2, 4.4){{\footnotesize $ 2 $}}
\put(1.6, 4.4){{\footnotesize $ 3 $}}
\put(2.45, 5.3){{\footnotesize $ 1 $}}
\put(3.5,3.0){(a)}
\multiput(1.5,1.0)(2,0){3}{\oval(1.1,0.2)}
\multiput(0.5,2.0)(2,0){4}{\oval(1.1,0.2)}
\multiput(0.0,1.5)(0,2){1}{\oval(0.2,1.1)}
\multiput(1.0,1.5)(0,2){1}{\oval(0.2,1.1)}
\multiput(2.0,1.5)(0,2){1}{\oval(0.2,1.1)}
\multiput(3.0,1.5)(0,2){1}{\oval(0.2,1.1)}
\multiput(4.0,1.5)(0,2){1}{\oval(0.2,1.1)}
\multiput(5.0,1.5)(0,2){1}{\oval(0.2,1.1)}
\multiput(6.0,1.5)(0,2){1}{\oval(0.2,1.1)}
\multiput(7.0,1.5)(0,2){1}{\oval(0.2,1.1)}
\put(3.2, 1.4){{\footnotesize $ 2 $}}
\put(2.45, 2.3){{\footnotesize $ 1 $}}
\put(3.5,0.0){(b)}

\end{picture}
\end{center}
\renewcommand{\baselinestretch}{0.8}
\caption[]{{Two arrangements of singlets in a two-dimensional array. }}
\end{figure}
At $J_{\perp}=0$
these configurations are degenerate. As was the case for the ladder (previous section),
to calculate ground state energy correction due to $J_{\perp}$
it is convenient to use the localized triplet representation for virtual excitations.
The energy of the localized triplet is $J$, and
the amplitude of two triplets created at the two singlet bonds coupled via 
$J_{\perp}$ is $J_{\perp}/4$, see e.g. Ref.\cite{Kotov}. Therefore, the second order 
correction to the
state Fig. 13a is described by the process $|0\rangle \to |12\rangle \to |0\rangle$
and is equal to $3(J_{\perp}/4)^2/(-2J)$. Here $|12\rangle$ denotes the state with triplets
excited at bonds $1$ and $2$ (see Fig. 13a), the coefficient, $3$, is number of
triplet polarizations. However, a similar contribution
exists for the configuration in Fig 13b and, therefore, the configurations Fig13a and Fig13b remain degenerate to second order in $J_{\perp}$.
To consider the next order we have to recall that the localized triplet can hop
between two bonds coupled via $J_{\perp}$. The hopping amplitude is also $J_{\perp}/4$,
see e.g. Ref.\cite{Kotov}. Additional corrections for Fig. 13a arise due to the possibility of hopping around the square (closed loop trajectory), 
whereas, in Fig. 13b, there is no such 
possibility. The correction is due to the following sequences of
 virtual states (see Fig. 13a):
\begin{eqnarray}
\label{sec}
|0\rangle &\to& |12\rangle \to |32\rangle \to |42\rangle \to |0\rangle,\\ 
|0\rangle &\to& |12\rangle \to |14\rangle \to |13\rangle \to |0\rangle,\nonumber\\ 
|0\rangle &\to& |12\rangle \to |32\rangle \to |34\rangle \to |0\rangle,\nonumber\\ 
|0\rangle &\to& |12\rangle \to |14\rangle \to |34\rangle \to |0\rangle.\nonumber\\
|0\rangle &\to& |12\rangle \to |1234\rangle \to |24\rangle \to |0\rangle.\nonumber\\
|0\rangle &\to& |12\rangle \to |1234\rangle \to |13\rangle \to |0\rangle.\nonumber 
\end{eqnarray}
The corresponding energy correction per one ``dot'' (one vertical $J_{\perp}$ 
cross-link) is then
\begin{equation}
\label{sec1}
E_{13a}-E_{13b}= -{15 \over 8}{{(J_{\perp}/4)^4}\over{J^3}}.
\end{equation}

We have thus shown that the system prefers the formation of ``squares''.
However, even in this case there is a pattern shown in Fig. 14 which, 
in the considered order ($J_{\perp}^4$), remains degenerate with that in Fig. 12.%
\begin{figure}
\begin{center}
\begin{picture}(7,4)
\multiput(1.5,0.0)(2,0){3}{\oval(1.1,0.2)}
\multiput(1.5,1.0)(2,0){3}{\oval(1.1,0.2)}
\multiput(0.5,2.0)(2,0){3}{\oval(1.1,0.2)}
\multiput(0.5,3.0)(2,0){3}{\oval(1.1,0.2)}
\multiput(1.0,0.5)(1,0){6}{\oval(0.2,1.1)}
\multiput(0.0,2.5)(1,0){6}{\oval(0.2,1.1)}
\end{picture}
\end{center}
\renewcommand{\baselinestretch}{0.8}
\caption[]{{Another arrangement of singlets in a two-dimensional array. }}
\end{figure}
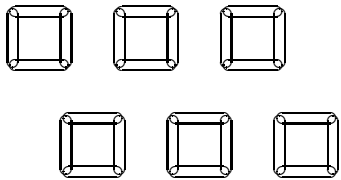
The degeneracy is lifted only in the $J_{\perp}^6$ term. To see this we first must recall
that the localized triplet can create an additional triplet at a nearby bond
along the MG chain: $t_{n\beta}^{\dag} \to t_{n\sigma}^{\dag}t_{n+1,\gamma}^{\dag}$
($t_{n\beta}^{\dag}$ is the localized triplet creation operator at the bond $n$,
$\beta$ is the triplet polarization). The amplitude of this process is
$\frac{1}{4}Jie_{\beta\sigma\gamma}$, see Ref. \cite {Shev}. Keeping this in mind, one can see
that the energy difference between Fig. 15a (building block of Fig. 12) and
Fig. 15b (building block of Fig. 14) is due to closed loop trajectories of the type
(for notations, see Fig. 15)
\begin{equation}
\label{sec2}
|0\rangle \to |12\rangle \to |125\rangle \to |126\rangle \to |127\rangle \to
|327\rangle \to |427\rangle \to |42\rangle \to |0\rangle.
\end{equation}
These trajectories exist for the configuration in Fig. 15a but do not exist 
for Fig. 15b. 
\begin{figure}
\begin{center}
\begin{picture}(7,6)
\multiput(1.0,4.5)(2,0){2}{\oval(1.1,0.2)}
\multiput(1.0,5.5)(2,0){2}{\oval(1.1,0.2)}
\multiput(0.5,5.0)(1,0){4}{\oval(0.2,1.1)}
\put(0.95,5.7){\footnotesize{1}}
\put(0.95,4.1){\footnotesize{4}}
\put(0.1,4.9){\footnotesize{3}}
\put(1.7,4.9){\footnotesize{2}}
\put(2.1,4.9){\footnotesize{6}}
\put(3.7,4.9){\footnotesize{8}}
\put(2.95,5.7){\footnotesize{5}}
\put(2.95,4.1){\footnotesize{7}}
\put(1.8,3.5){(a)}
\multiput(1.0,1.5)(2,0){2}{\oval(1.1,0.2)}
\put(1.0,2.5){\oval(1.1,0.2)}
\put(3.0,0.5){\oval(1.1,0.2)}
\multiput(0.5,2.0)(1,0){2}{\oval(0.2,1.1)}
\multiput(2.5,1.0)(1,0){2}{\oval(0.2,1.1)}
\put(1.8,0.0){(b)}
\end{picture}
\end{center}
\renewcommand{\baselinestretch}{0.8}
\caption[]{{(a) Building block for Fig. 12a. (b) Building block for 
Fig 14. }}
\end{figure}
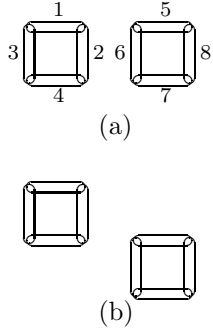
Calculation of the matrix element corresponding to the 
trajectory (\ref{sec2}) and counting all the trajectories we find the energy difference,
\begin{equation}
\label{edif}
E_{15a}-E_{15b}=-{1\over144}{{(J_{\perp}/4)^6}\over{J^5}}.
\end{equation}
Thus, We have shown that the configuration presented in Fig. 12 is the true 
ground state of the Hamiltonian (\ref{hamAB}). Note that the parameter of expansion 
in Eqs. 
(\ref{sec1}) and (\ref{edif}) is $J_{\perp}/4J$. Note also that the coefficient in
(\ref{edif}) is anomalously small meaning that the system is quite soft with
respect to the dislocations shown in Fig. 14.

The lowest energy excitation consists of two spinons, see Fig. 16. 
\begin{figure}
\begin{center}
\begin{picture}(10,3)
\multiput(0.5,0.5)(2,0){5}{\oval(1.1,0.2)}
\put(0.5,1.5){\oval(1.1,0.2)}
\multiput(3.5,1.5)(2,0){2}{\oval(1.1,0.2)}
\put(2.0,1.5){\vector(0,1){0.7}}
\put(7.0,1.5){\vector(0,1){0.7}}
\put(8.5,1.5){\oval(1.1,0.2)}
\multiput(0.0,1.0)(1,0){10}{\oval(0.2,1.1)}
\put(9.5,0.4){A}
\put(9.5,1.4){A}
\end{picture}
\end{center}
\renewcommand{\baselinestretch}{0.8}
\caption[]{{ Two spinons along one of the A-chains in a 
two-dimensional array. }}
\end{figure}
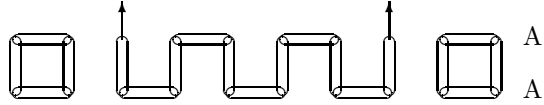
Here the spinons
are created on the upper A-chain and 
all other spins remains paired. According to (\ref{sec1}), the 
tension of the confining string is
\begin{equation}
\label{tt}
T={15 \over 4} {{(J_{\perp}/4)^4}\over{J^3}}
\end{equation}
and hence the dynamics are described by Eqs. (\ref{schr}), (\ref{sp}) and (\ref{psi})
with substitutions
\begin{eqnarray}
\label{subsAB}
&&{{3\delta J}\over{4}} \to {15 \over 4} {{(J_{\perp}/4)^4}\over{J^3}},\\
&&\xi \to \left({8 \over 15}\right)^{1/3}\left({{4J}\over{J_{\perp}}}\right)^{4/3}.
\nonumber
\end{eqnarray}
We now intend to consider the spinons tunneling from one chain to another.
This is simpler to do for triplet excitations and therefore we concentrate
only on this case. 
Taking account of (\ref{psi}) and (\ref{subsAB}),
the wave functions of triplets built on the A- and B-chains are
\begin{eqnarray}
\label{psiAA}
\Psi_A({\bf r}_1,{\bf r}_2)&=&
{{e^{i{\bf KR}}}\over{\sqrt{N}}}
\frac{Ai\left((x+ 1)/\xi + z_n\right)}{\sqrt{\xi}\left|Ai^{\prime}(z_n)\right|}
\delta_{y_1,y_2}, \\
\Psi_B({\bf r}_1,{\bf r}_2)&=& 
{{e^{i{\bf KR}}}\over{\sqrt{N}}}
\frac{Ai\left((y+ 1)/\xi + z_n\right)}{\sqrt{\xi}\left|Ai^{\prime}(z_n)\right|}
\delta_{x_1,x_2}, \nonumber
\end{eqnarray}
where ${\bf R}=({\bf r}_1+{\bf r}_2)/2$, ${\bf r}={\bf r}_2-{\bf r}_1$.
Eqs. (\ref{psiAA}) do not make account of the tunneling between the chains.
The mechanism of the tunneling is similar to that of the ladder (see Fig. 9) - 
the spinons approach one another to within one lattice space and then the
perturbation, $J_{\perp}{\bf S}_A{\bf S}_B$, in the Hamiltonian (\ref{hamAB}) swaps 
the triplet from the A-chain to the B-chain (and vice versa). 
Calculation of the tunneling amplitude gives
\begin{equation}
\label{tun1}
\tau=J_{\perp}\cos {{K_x}\over{2}} \cos {{K_y}\over{2}} {4\over{\xi^3}}
= 30J \cos {{K_x}\over{2}} \cos {{K_y}\over{2}} \left({{J_{\perp}/4}\over{J}}\right)^5.
\end{equation}
The triplet dispersion is given by diagonalizing the $2 \times 2$ matrix
\begin{eqnarray}
\label{ma}
\pmatrix{ \epsilon_A & \tau \cr \tau & \epsilon_B } ,
\end{eqnarray}
where
\begin{eqnarray}
\label{EAB}
\epsilon_A({\bf K})&=&2\Delta-2z_n J \left({3\over2}\right)^{2/3} 
\left({{J_{\perp}}\over{4J}}\right)^{8/3}-3{{\left(J_{\perp}/4\right)^4}\over{J^3}}
+{{K_x^2}\over{2M}}  \\
\epsilon_B({\bf K})&=&2\Delta-2z_n J \left({3\over2}\right)^{2/3} 
\left({{J_{\perp}}\over{4J}}\right)^{8/3}-3{{\left(J_{\perp}/4\right)^4}\over{J^3}}
+{{K_y^2}\over{2M}}. 
\end{eqnarray}
Here $M=2m=1/J$ is total mass of triplet excitation.
Diagonalization of (\ref{ma}) gives
\begin{equation}
\label{disp1}
\epsilon={{\epsilon_A+\epsilon_B}\over 2}\pm 
\sqrt{\left({{\epsilon_A-\epsilon_B}\over 2}\right)^2+\tau^2}.
\end{equation}
At very small momenta, ${\bf K}^2/2M \ll \tau \propto J_{\perp}^5$, rotational
invariance is restored and we have two distinct triplet excitations with a
quadratic dispersion
\begin{equation}
\label{disp2}
\epsilon \approx {{\epsilon_A+\epsilon_B}\over 2}\pm \tau =const +{{{\bf K}^2}\over{2M}}
\pm \tau.
\end{equation}

\section{Conclusions}
In the present work we have considered several models in the regime close to 
the deconfinement of spinons. These systems have complex spectra which consist of multiple singlet and triplet states. 

For the one dimensional  $J_1-J_2-\delta$ model with
$J_2/J_1 \approx 0.5$ and $\delta \ll 1$ we have calculated the splitting between
the singlet and triplet states as well as the spectral weight of triplet excitations.
The splitting is in a good agreement with available numerical data. On the other hand, the spectral weight is not consistent with results of series expansion. This may be explained by the very slow convergence of the series for the system which has a dense spectrum. It would be very interesting to study this problem using different numerical methods.

For the $J_1-J_2-J_{\perp}$ ladder ($J_{\perp} \to 0$) we found that the 
interval between levels is
$\Delta E\propto J_{\perp}^{4/3}$. Each level has a fine structure: it consists
of two triplet and two singlet states. The triplet-singlet splitting is 
$\propto J_{\perp}^2$. Splitting between the triplets is $\propto J_{\perp}^3$,
which is due to tunneling between the legs. The typical size of the states
under consideration is $\xi \propto J_{\perp}^{-2/3}$. This is the radius of
confinement.

Similar results are obtained for the two dimensional array of coupled $J_1-J_2$
chains (two dimensional $J_1-J_2-J_{\perp}$ model as $J_{\perp} \to 0$).
The energy interval between the levels is $\Delta E\propto J_{\perp}^{8/3}$.
Fine structure of the level consists of two singlets and two triplets.
The triplet-singlet splitting is $\propto J_{\perp}^4$. Splitting between the 
triplets due to tunneling is $\propto J_{\perp}^5$.
The confinement radius is $\xi \propto J_{\perp}^{-4/3}$.
As does any lattice model, this model violates rotational invariance. However, in the low energy
sector ($K \ll (J_{\perp}/J)^{5/2}$), rotational invariance is restored.

\section*{Acknowledgment}
We are very thankful to V. N. Kotov and R. R. P. Singh for stimulating discussions.
We are also very thankful to E. S. S{\o}resen for very helpful comments
and providing data prior to publication. It is a pleasure to
acknowledge discussions with Z. Weihong.


\begin{thebibliography}{10}
\bibitem{norm} We differ in normalization with Ref. \cite{Zheng} by a factor of 2. For example, in
our normalization, at $\delta=1$ and $K=\pi$, the spectral weight is equal to unity.
\bibitem{cor} Calculation of $\Delta E$ is similar to that performed in Refs.
\cite{Sushkov,Kotov}.
\end{thebibliography}

\end{document}